\documentclass[%
 reprint,nofootinbib,
superscriptaddress,
 amsmath,amssymb,aps,
 prb,
]{revtex4-2}

\usepackage{graphicx}
\usepackage{dcolumn}
\usepackage{bm}
\usepackage{xcolor}
\usepackage{amsmath,amssymb}
\usepackage{mathtools}
\usepackage{ulem}
\usepackage{cancel}

\usepackage{xcolor}

\begin{document}

\title{Topological edge states in curved zigzag superlattices in nonlinear exciton-polaritons}

\author{Jing Wang$^{\dagger}$}
\affiliation{School of Integrated Circuits, Guangdong University of Technology, Guangzhou 510006, China}

\author{Tobias Schneider$^{\dagger}$}
\affiliation{Department of Physics and Center for Optoelectronics and Photonics Paderborn (CeOPP), Paderborn University, 33098 Paderborn, Germany}

\author{Wei Hu}
\affiliation{Guangdong Provincial Key Laboratory of Nanophotonic Functional Materials and Devices, South China Normal University, Guangzhou 510631, China}

\author{Stefan Schumacher}
\affiliation{Department of Physics and Center for Optoelectronics and Photonics Paderborn (CeOPP), Paderborn University, 33098 Paderborn, Germany}
\affiliation{Institute for Photonic Quantum Systems (PhoQS), Paderborn University, 33098 Paderborn, Germany}
\affiliation{Wyant College of Optical Sciences, University of Arizona, Tucson, Arizona 85721, USA}

\author{Xuekai Ma}
\email{xuekai.ma@gmail.com}
\affiliation{Department of Physics and Center for Optoelectronics and Photonics Paderborn (CeOPP), Paderborn University, 33098 Paderborn, Germany}

\begin{abstract}
Zigzag chains allow for the formation of topological edge states. Several distinct chain architectures have been developed for this purpose. Here, we report a zigzag superlattice, containing two staggered sub-lattices, that supports multiple edge states, including higher-order modes. In such lattices, the intra- and intercell coupling is imbalanced by the tunneling effect of the eigenstates or deformation of the higher-order modes. We demonstrate that by arranging the zigzag superlattice into a curved shape, some of the edge states transition into bulk states as the curvature of the lattice increases, while some bulk states become more localized towards edge states. The reason is that a curved superlattice strengthens the intra-lattice coupling of the inner sub-lattice due to the separation reduction of the potential wells. 
We also show that some bulk states at larger curvatures can be transformed into edge states by a repulsive nonlinearity, which also enables the coexistence of different edge states. As a specific and ideal platform for realizing such topological superlattices we explore exciton-polaritons in semiconductor microcavities with their strong nonlinearity and possibility for optical excitation and control. Our work introduces an additional dimension for the design of complex topological lattices and functional photonic devices.
\end{abstract}

\maketitle
\def\thefootnote{$\dagger$}\footnotetext{These authors contributed equally to this work}\def\thefootnote{\arabic{footnote}}

\section{Introduction}
One-dimensional (1D) zigzag chains are an intriguing structure that can support topological edge states, which can be described by the Su–Schrieffer–Heeger (SSH) model~\cite{PhysRevLett.42.1698}. The SSH chains are sought-after, ranging from atoms to photonics, not only because of the structural simplicity that can be realized in diverse physical systems, but also because of the exotic topological properties that can be used to develop functional devices. For example, the SSH model shows sharp state transitions of polarons~\cite{PhysRevLett.105.266605} and formation of localized topological solitons in the molecule trans-polyacetylene~\cite{meier2016observation}. Robust quantum state transfer can be realized in superconducting qubit chains~\cite{PhysRevA.98.012331}, and two-photon quantum walks have been observed in a photonic SSH lattice~\cite{klauck2020photonic}. The SSH model can be extended to non-Hermitian systems, leading to robust photonic zero modes~\cite{pan2018photonic} and the non-Hermitian skin effect~\cite{PhysRevResearch.1.023013}. Topologically protected lasing in an exciton-polariton SSH chain in a semiconductor microcavity has recently triggered significant interest~\cite{st2017lasing}.

Exciton-polaritons are light-matter quasiparticles that form in semiconductor microcavities via the strong coupling of photons and excitons. The hybrid nature enables optical accessibility and shows prominent nonlinearity, also facilitating polariton condensation~\cite{byrnes2014exciton}. A larger exciton binding energy even promises room-temperature operation as shown in perovskites~\cite{su2020observation} and organic materials~\cite{plumhof2014room,daskalakis2014nonlinear}. Room temperature polaritonics is a rapidly growing field, benefiting from not only the simplified measurements and operation but also the fabrication of micro- and nanostructures on top of the samples, which accelerates the development of functional polariton devices. One of the aims of the microstructure fabrication and design on the samples is to study topological insulators, with the goal to develop topologically protected lasers, for instance. Popular topological lattices include honeycomb lattices~\cite{klembt2018exciton,milicevic2015edge}, Lieb lattices~\cite{PhysRevLett.120.097401,PhysRevB.97.081103}, kagome lattices~\cite{gulevich2017exploring,PhysRevB.94.115437}, disclination lattices~\cite{jin2025perovskite,sabour2025polariton}, and SSH chains~\cite{PhysRevLett.116.046402,st2017lasing,su2021optical,pieczarka2021topological,harder2021coherent,georgakilas2025situ,zhao2025observation}. The topological insulator states in these structures can be enriched by nonlinearity and nonequilibrium, such as bistability~\cite{PhysRevLett.119.253904,Ma:20} and modulational instability~\cite{Kartashov:16,PhysRevA.99.053836} of edge currents, topologically protected solitons~\cite{Kartashov:16,gulevich2017exploring}, and gap solitons~\cite{pernet2022gap}. The occurrence of topological edge states in SSH chains requires a larger intercell coupling than the intracell coupling. This can be fulfilled for example by controlling the separation of the potential wells in 1D chains~\cite{georgakilas2025situ,zhao2025observation}, in virtue of directional dipole modes in zigzag chains~\cite{st2017lasing,pieczarka2021topological,harder2021coherent}, by means of TE-TM splitting and polarization properties~\cite{PhysRevLett.116.046402,su2021optical}, or utilizing coupled wave lattices which provide another degree of freedom and allow multistable edge states~\cite{schneider2024topological}. A 1D SSH chain can be extended to contain more periods in the chain direction to support edge states~\cite{PhysRevA.98.043838,zhang2021topological} or be extended along the perpendicular direction to a two-dimensional (2D) SSH lattice to support edge and corner states~\cite{PhysRevB.98.205147,PhysRevLett.122.233903,wu2023higher,bennenhei2024organic}.

\begin{figure*}[t]
\centering
{\includegraphics[width=\linewidth]{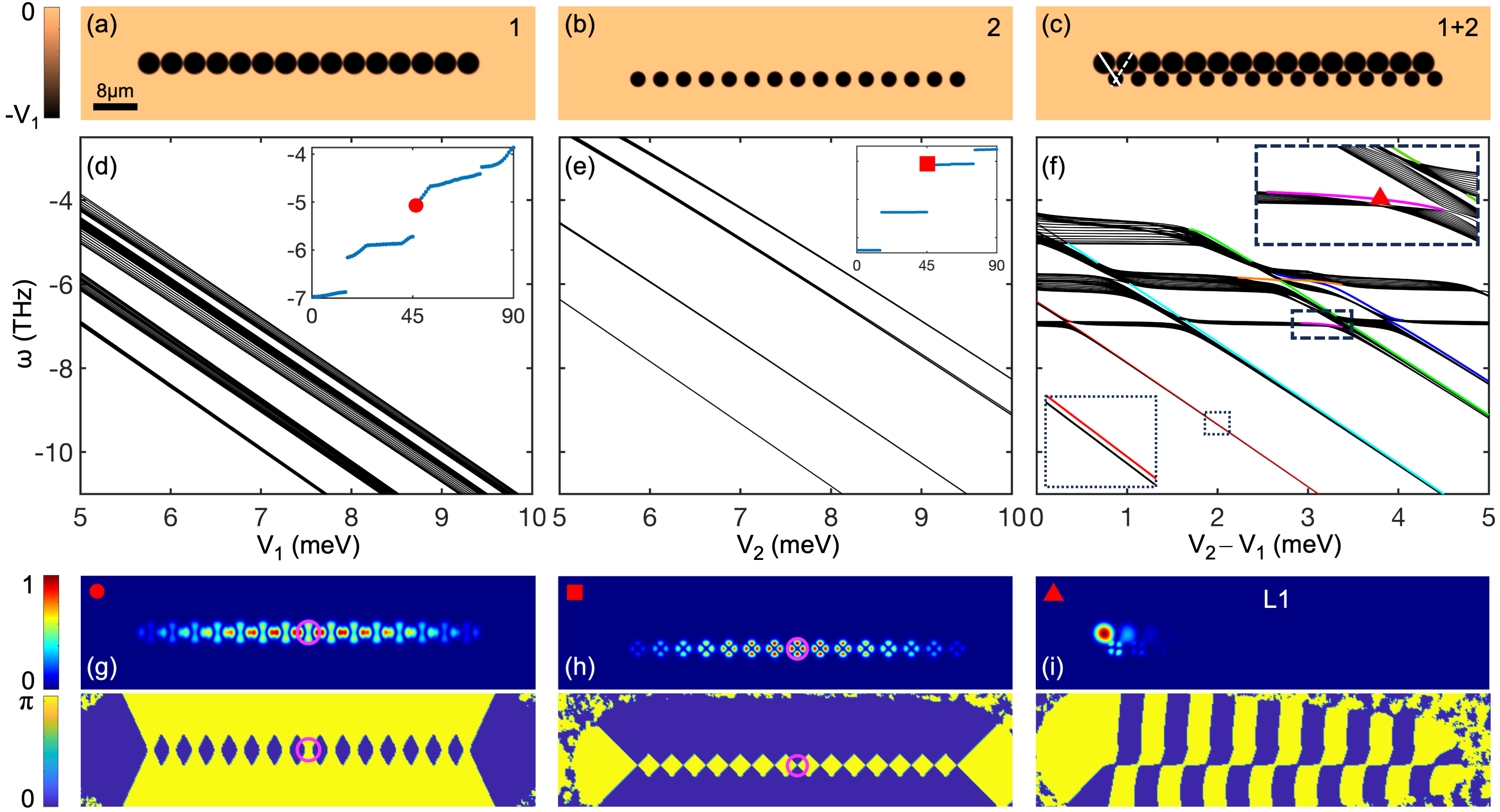}}
\caption{\textbf{Lattice structures and eigenstates.} (a) Distribution of a potential chain, Lattice-1, containing 15 identical potential wells with a diameter of 4~$\mu$m and a separation, center to center, of 4~$\mu$m. (b) A similar chain with a smaller well diameter of 2.8~$\mu$m, Lattice-2. (c) A superlattice composed of staggered Lattice-1 and Lattice-2 with the separation of the nearest larger and smaller potential wells of 3.4~$\mu$m, forming a perfect zigzag shape. The solid and dashed lines indicate the intracell and intercell coupling directions, respectively. (d,e) Dependence of the first four energy bands on the potential depth of Lattice-1 and Lattice-2, respectively. The inset in (d) shows the first 90 eigenvalues at the potential depth of $V_1$=5~meV, and the inset in (e) shows the first 90 eigenvalues at the potential depth of $V_2$=8~meV. (f) Dependence of the energy bands of the zigzag superlattice in (c) on the difference of the potential depths with $V_1$=5~meV fixed. Black lines represent bulk states, while colored lines represent edge states. The insets are the enlarged view of the corresponding dashed boxes. (g-i) Amplitude (upper) and phase (lower) profiles of the eigenstates marked in (d-f). The circles in (g,h) indicate an individual potential well in each lattice.}
\label{fig:1}
\end{figure*}

In the present work, we report another mechanism to realize topological edge states based on a zigzag superlattice, containing two sub-lattices. When the two sub-lattices have different sizes and depths of the potential wells, two types of eigenstates from different energy bands can coexist in the superlattice at a specific energy, and their coupling is able to significantly imbalance the intra- and intercell coupling of the zigzag superlattice, which is essential for realizing edge states in either of the sub-lattices. The relative role of intra- and intercell coupling can be further tuned by bending the zigzag chain into an arc or a circle with finite curvature. In this scenario, some edge states permeate into the bulk potential wells, which can be prevented by nonlinearity. Under resonant excitation, the coexistence of different edge states at both edges can be achieved. The curvature-induced separation reduction of the potential wells in one sub-lattice is different from the direct variation of the potential separations in a sub-lattice, as the finite curvature preserves the perfect zigzag shape, which is important for the occurrence of edge states and uniform inter-lattice coupling. In addition, the curvature leads to an overall less elongated structure which may benefits integration and engineering of complex topological structures.

\begin{figure*}[t]
\centering
{\includegraphics[width=\linewidth]{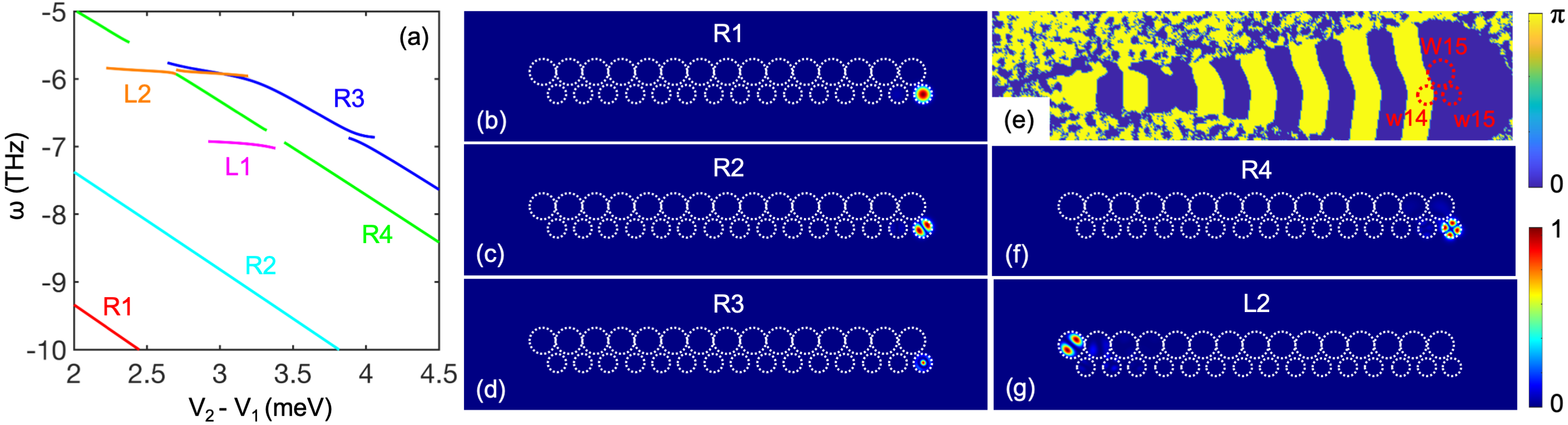}}
\caption{\textbf{Edge eigenstates in standard zigzag superlattices.} (a) Dependence of edge states on the difference of the potential depths of the two sub-lattices with $V_1$=5~meV fixed, extracted from Fig.~\ref{fig:1}(c). Amplitude distribution of a (b) R1, (c) R2, (d) R3, (f) R4, and (g) L2 edge state with the lattice structure marked on top. (e) Phase of the R1 edge state in (b) with the last three potential wells marked on top.}
\label{fig:2}
\end{figure*}

\section{Theoretical Model}
The dynamics of polariton condensates under resonant excitation can be described by a generalized driven-dissipative Gross-Pitaevskii (GP) equation, which is also known as a dissipative nonlinear Schr{\"o}dinger equation, i.e., 
\begin{equation}
\begin{aligned}
\label{GP_psi}
    i\hbar\frac{\partial \psi(\textbf{r},t)}{\partial t} =&\left[-\frac{\hbar^2}{2m}\nabla^2-i\hbar\frac{\gamma}{2}+g|\psi(\textbf{r},t)|^2+V(\textbf{r})\right]\psi(\textbf{r},t) \\
    \\
    &+E(\textbf{r},t)\,.
\end{aligned}
\end{equation}
Here, $\psi(\textbf{r},t)$ is the polariton wavefunction. $m=10^{-4}m_\text{e}$ is the effective polariton mass with $m_\text{e}$ being the free electron mass. $\gamma=0.005$~ps$^{-1}$ is the loss rate, and the resonant driving is realized by a coherent pump $E(\textbf{r},t)$. The repulsive polariton-polariton interaction is denoted by $g=2~\mu$eV~$\mu$m$^{2}$, i.e., the nonlinearity coefficient. The potential $V(\textbf{r})$ contains the spatial distribution of the lattices. In this work, two types of lattices are defined: Lattice-1 contains 15 identical round potential wells with a diameter of 4~$\mu$m, depth of $V_1$, and lattice constant of 4~$\mu$m [Fig.~\ref{fig:1}(a)]; Lattice-2 also contains 15 identical round potential wells with a smaller diameter of 2.8~$\mu$m, depth of $V_2$, and a lattice constant of 4~$\mu$m [Fig.~\ref{fig:1}(b)]. A superlattice with a perfect zigzag shape can be formed by placing Lattice-1 and Lattice-2 next to each other with contact and spatial stagger; see Fig.~\ref{fig:1}(c). 

\section{Edge States in Standard Zigzag Superlattices}
Linear eigenstates in such lattices can be obtained by solving the linear eigenvalue problem by neglecting the loss, gain, and nonlinearity in Eq.~\eqref{GP_psi} and assuming the wavefunction to have the form $\Psi(\textbf{r},t)=\Psi(\textbf{r})e^{-i{\omega}t}$, i.e., 
\begin{equation}
\label{eigenvalue}
    \hbar\omega{\psi(\textbf{r})} =\left[-\frac{\hbar^2}{2m}\nabla^2+V(\textbf{r})\right]\psi(\textbf{r}).
\end{equation}

In Lattice-1 in Fig.~\ref{fig:1}(a), the potential wells have a larger diameter and touch each other, resulting in their eigenstates being distributed in the frequency domain, as shown in Fig.~\ref{fig:1}(d), because of the larger intra-lattice coupling. In contrast, the energy bands are clearly separated in Lattice-2; see Figs.~\ref{fig:1}(b,e), as the potential wells are well separated. Deeper potential wells drag all the energy bands to lower frequencies/energies [Figs.~\ref{fig:1}(d,e)]. In the superlattice composed of Lattice-1 and Lattice-2 [Fig.~\ref{fig:1}(c)], if the potential depth of Lattice-1 is fixed at $V_1$=5~meV and the potential depth of Lattice-2, $V_2$, is gradually varied, a more complex energy band structure is produced as shown in Fig.~\ref{fig:1}(f), in which several edge states marked with colored lines in the gaps can be recognized with an example plotted in Fig.~\ref{fig:1}(i). It is known from the SSH model that in zigzag chains, topological edge states occur when the intercell coupling is stronger than the intracell coupling~\cite{PhysRevLett.42.1698}. The L1 edge state in Fig.~\ref{fig:1}(i) also follows this rule. From Figs.~\ref{fig:1}(g,h), one can see that the enhanced intra-lattice coupling can generate deformed higher-order modes (a quadrupole here), which in independent potential wells remain non-deformed, cf. the amplitude and phase profiles of the states within the circles. Likewise, the quadrupole modes in Lattice-2 start to deform when it approaches Lattice-1 to form the zigzag chain, and the deformation of the quadrupole modes can lead to imbalanced intra- and intercell couplings, which enables the formation of the L1 edge state. It is worth noting that due to the difference of the potential depths in Lattice-1 and Lattice-2, the eigenfrequency of the L1 edge state in Fig.~\ref{fig:1}(i) corresponds to the fundamental mode of Lattice-1, see Fig.~\ref{fig:1}(f), i.e., it is a fundamental edge state.

The edge states in Fig.~\ref{fig:1}(f) are hard to distinguish because of the larger energy range. To give a better overview, they are extracted and presented in Fig.~\ref{fig:2}(a), from which five additional edge states, R1, R2, R3, R4, and L2 states, can be clearly seen. The amplitude and phase distributions of these edge states are shown in Figs.~\ref{fig:2}(b-g), including the right edge states, R1, R2, and R3, in Lattice-2 and the left edge state, L2, in Lattice-1. Note that there are more edge states of much higher-order modes in this configuration, which are neglected, as in this work, we only consider the first four energy bands, i.e., the fundamental mode, dipole mode, quadrupole mode, and a Laguerre-Gaussian mode. 

It is known that dipole edge states [Figs.~\ref{fig:2}(c,g)] appear when the longitudinal orientation aligns with the intercell coupling direction; see the dashed line in Fig.~\ref{fig:1}(c)~\cite{st2017lasing}. This can also be applied to a Laguerre-Gaussian mode [Fig.~\ref{fig:2}(d), which becomes a tripole mode when squeezed] and quadrupole Fig.~\ref{fig:2}(f) modes as long as they are deformed along the intercell coupling direction, as discussed in Fig.~\ref{fig:1}(i). However, the fundamental modes are hard to squeeze, raising the question of why the R1 edge state exists in Fig.~\ref{fig:2}(b). To illustrate it, the phase profile of the R1 edge state is provided in Fig.~\ref{fig:2}(e), in which from the last three marked potential wells one can see that the two sub-lattice ends, W15 and w15, share the same phase with a $\pi$ phase jump between their phase and the phase in w14. It shows in Fig.~\ref{fig:2}(a) that the energy of the fundamental mode ($\sim6$~meV) is below the potential depth (5~meV) of Lattice-1, so that there are no eigenstates allowed in Lattice-1 at this energy level, that is, the inter-lattice coupling should in principle be zero. Nevertheless, the fundamental modes in the w14 and w15 wells in Lattice-2 can tunnel into W15. If the tunneled tail in W15 is from w14, i.e., they share the same phase, there is no coupling between W15 and w14, because they belong to the same state, but a weak coupling between W15 and w15. In this case, the intercell coupling is weaker than the intracell coupling, making it impossible to form edge states. Instead, the intercell coupling becomes stronger than the intracell coupling when the tunneled tail in W15 originates from w15, leading to the occurrence of the fundamental edge state presented in Fig.~\ref{fig:2}(b). This mechanism also applies to the other right edge states in Fig.~\ref{fig:2}, when their eigenenergies are below the first energy band of Lattice-1. When their eigenenergies are larger than that, the mode deformation as discussed in Fig.~\ref{fig:1}(i) plays a crucial role for the occurrence of edge states. Therefore, the formation mechanism of the dipole edge states in Figs.~\ref{fig:2}(c,g) is different from that studied in previous works~\cite{st2017lasing,pieczarka2021topological,harder2021coherent}.

\section{Edge States in Curved Zigzag Superlattices}
\begin{figure*}[t]
\centering
{\includegraphics[width=1\linewidth]{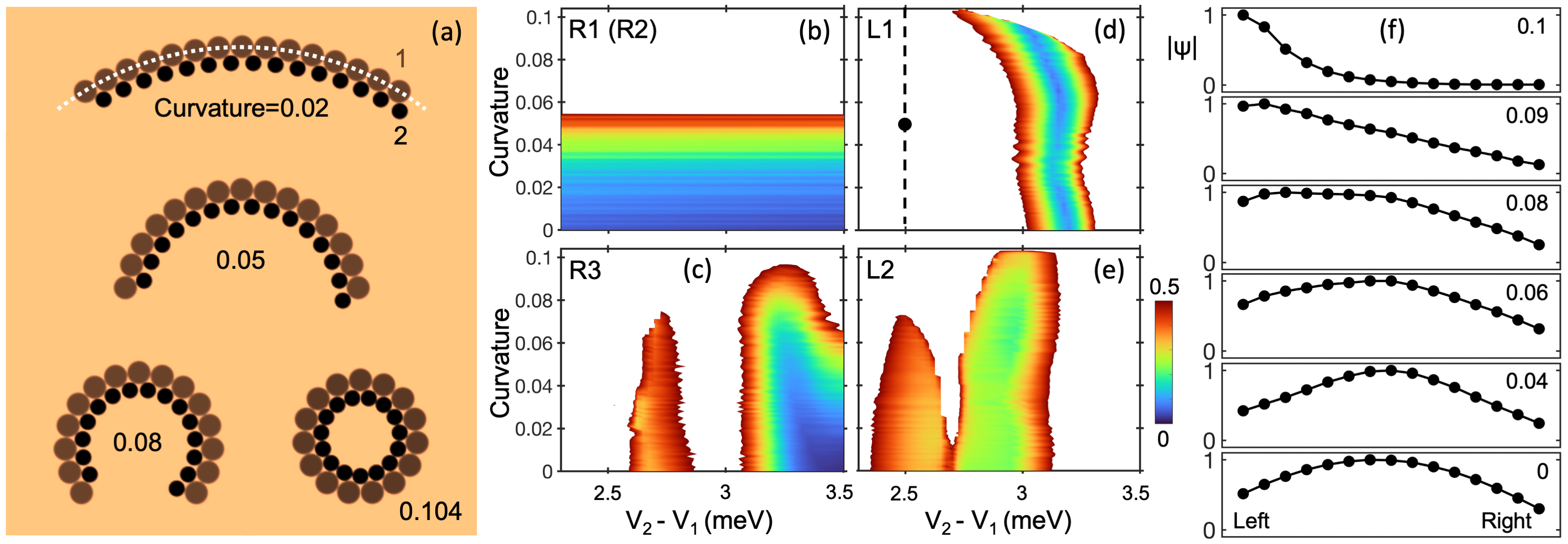}}
\caption{\textbf{Edge states in curved superlattices.} (a) Examples of differently curved chains with $V_1$=5~meV and $V_2$=7~meV. Existence areas of the (b) R1 and R2, (c) R3, (d) L1,  and (e) L2 edge states in relation to the potential depth difference (here $V_1$=5~meV) and the curvature of the chain. The colored areas represent edge states, while the white areas represent bulk states. The colorbar indicates the ratio of the peak amplitude in the second well from the edge (in the same sub-lattice) and the peak amplitude in the edge well, i.e., $|\Psi^\textup{Lattice-1(2)}_{\textup{edge}\pm1}|^\textup{max}/|\Psi^\textup{Lattice-1(2)}_{\textup{edge}}|^\textup{max}$. The smaller the ration, the more prominent the edge state. (f) Normalized peak amplitudes of all potential wells in Lattice-1, indicated by the dashed curve in (a), of the bulk states selected along the dashed line in (d) at different curvatures.}
\label{fig:3}
\end{figure*}

In 2D SSH chains, it has been demonstrated that the next-nearest-neighbor coupling, referring to the intra-lattice coupling in our case, determines the formation of corner states~\cite{PhysRevB.98.205147,PhysRevLett.122.233903}. In our work, we tune the intra-lattice coupling by bending the zigzag chain to reduce the separation of the potential wells in Lattice-2, but keep the separation of the potential wells in Lattice-1 fixed. Several examples of arched zigzag superlattices are shown in Fig.~\ref{fig:3}(a) with different curvatures. Based on the fixed number of potential wells and their sizes, the curvature of the superlattice chain ranges from 0 [a standard zigzag chain, in which the potential wells in Lattice-2 have the largest separation, as shown in Fig.~\ref{fig:1}(c)] to 0.104 (a closed circular chain, in which the potential wells in Lattice-2 touch each other). The curvature is defined as the reciprocal of the ring that is formed by the centers of the potential wells in Lattice-1, i.e., $1/R_{\textup{Lattice-1}}$; see the dashed line in Fig.~\ref{fig:3}(a). Apparently, no edge states are possible at the curvature of 0.104 as the chain is closed. 

To study the influence of the curvature on the edge states, in this work, the definition of the edge states satisfies the relation of $|\Psi^\textup{Lattice-1(2)}_{\textup{edge}\pm1}|^\textup{max}/|\Psi^\textup{Lattice-1(2)}_{\textup{edge}}|^\textup{max}\leq0.5$, that is, if the maximum amplitude of the state in the neighbor of the edge well of the same sub-lattice is smaller than or equal to half of the maximum amplitude of the state in the edge well, it is an edge state; otherwise it is regarded as a bulk state. Figures~\ref{fig:3}(b-e) show how this relation varies with the curvature and the potential depth difference. The colored areas represent the edge states, while the white areas are bulk states. The blue color means that the edge state is more distinct than that represented by the red color. It can be seen from Fig.~\ref{fig:3}(b) that the R1 and R2 edge states gradually seep into the bulk potential wells as the curvature of the superlattice increases and evolve into bulk states when the curvature is larger than 0.55, which is independent of the potential depth difference. For these two edge states, the curvature dependence is ascribed to the coupling change of the potential wells in Lattice-2, which affects the tunneling of the states from Lattice-2 to Lattice-1. In other words, a smaller separation of the potential wells w14 and w15 in Fig.~\ref{fig:2}(e) favors the tunneling of the states between them and hence prevents their tunneling into W15, which consequently rebalances the intra- and intercell coupling and extinguishes the edge states.

Normally, higher-order excited states are more easily influenced, but compared with R1 and R2 edge states, the R3 edge states are less affected by the curvature, as can be seen in Fig.~\ref{fig:3}(c). This is because within this potential depth difference and eigenfrequencies, there are already eigenstates in Lattice-1 to enhance the inter-lattice coupling. Similarly, as the intra-lattice coupling in Lattice-2 is increased with the curvature, the existence area of the R3 edge state reduces. At the curvature around 0.09, the two edge potential wells approach each other, so that a higher-order mode has a larger possibility to cross the barrier between them and build the connection of the two edges, leading to the disappearance of the edge states. The existence area of the R3 edge state is split on account of the influence of the L2 edge state, as their energies are very close to each other around $V_2-V_1$=3 meV [see Fig.~\ref{fig:2}(a)]. As a result, in this area, there appears a protuberant amplitude peak at the right edge of Lattice-1, the nearest neighbor of the R3 peak (see the density profiles in Fig.~\ref{fig:5} below). 

The left edge states in Lattice-1, however, show a distinct reaction to the curvature of the chain, which even expands the existence area of the L1 edge state, as shown in Fig.~\ref{fig:3}(d). Interestingly, the area of the blue color, representing prominent edge states, holds as the curvature varies, besides shifting slightly along with the potential depth difference. The cutting edge at larger curvatures is due to the redshift of the energy band just above the gap because of the increased $V_2$. For the same reason, the existence area of the L2 edge state is split into two parts, as presented in Fig.~\ref{fig:3}(e), by an energy band, which is different from the splitting of the R3 edge state in Fig.~\ref{fig:3}(c). As a consequence, in the left (right) area of the L2 edge state, the eigenstates in the inner Lattice-2 are quadrupole (tripole, see Fig.~\ref{fig:5}(e) below) modes. Also, the influence of the curvature on the existence area of the L2 edge state is insignificant; see the left boundary of the left area and the right boundary of the right area in Fig.~\ref{fig:3}(e).

\begin{figure}[t]
\centering
{\includegraphics[width=1\linewidth]{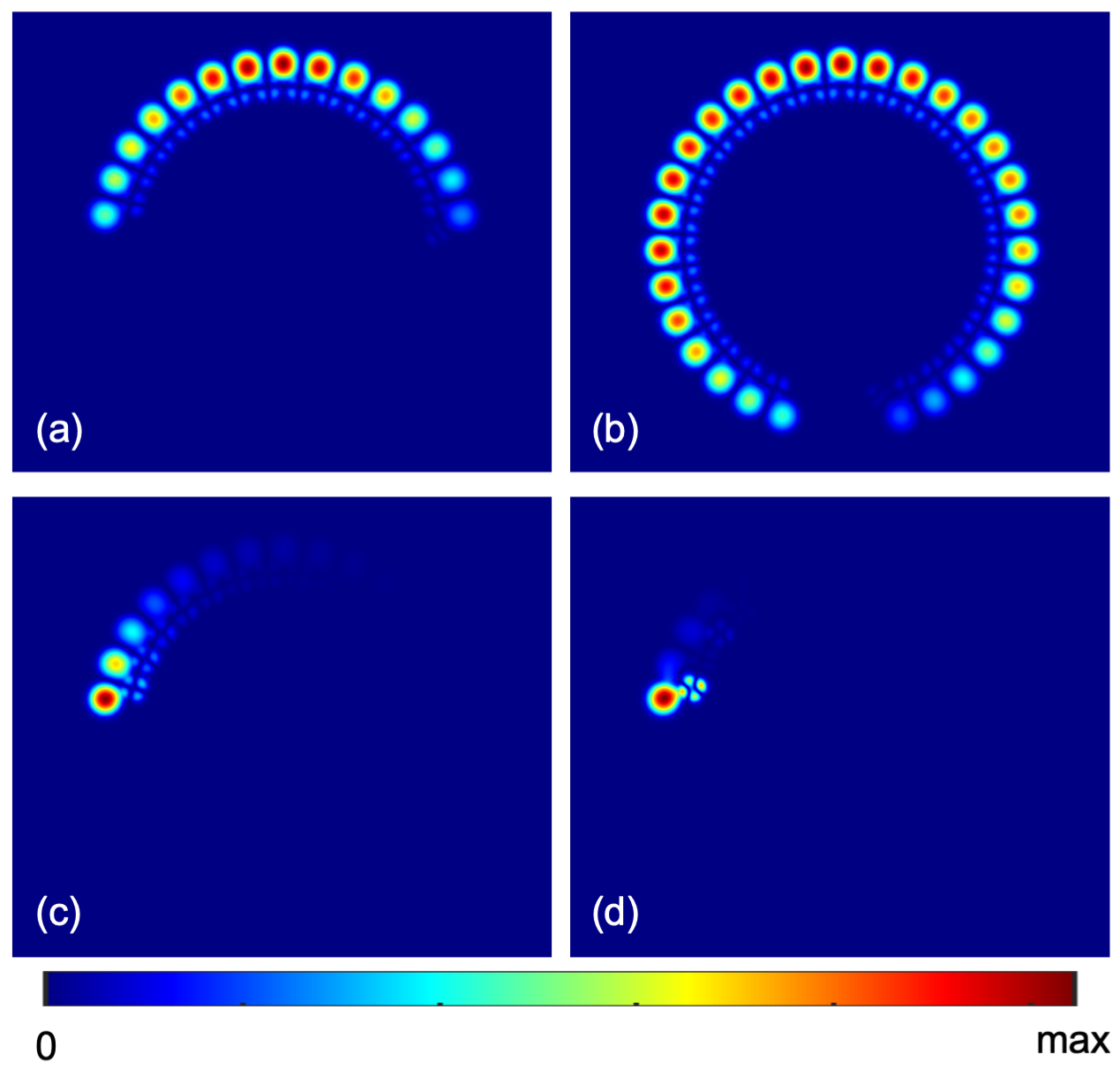}}
\caption{\textbf{Edge states enhanced by the intra-lattice coupling.} Amplitude distributions of (a) the state in a zigzag chain with the curvature of 0.05, 15 wells, $V_1$=5~meV, and $V_2$=7.5~meV, corresponding to the state marked by the black point in Fig.~\ref{fig:3}(d), (b) the state in a similar chain as in (a) but more, 31, potential wells, (c) the state in a similar chain as in (a) but a increased well diameter, 3~$\mu$m, of the potential wells in Lattice-2, and (d) the state in a similar chain as in (c) but a larger potential depth, $V_2$=7.75~meV, of the potential wells in Lattice-2.}
\label{fig:4}
\end{figure}

To understand how the existence area of the L1 edge state is being expanded by the curvature, we focus on the bulk state at $V_2-V_1$=2.5~meV [the dashed line in Fig.~\ref{fig:3}(d)] and extract the maximum amplitude of all potential wells in Lattice-1. The curvature-dependent results are shown in Fig.~\ref{fig:3}(f), from which one can see that a smaller curvature, 0.04 for instance, strengthens the amplitude in the middle wells. When the curvature is increased further, the peak amplitude starts to move gradually to the left edge, approaching the L1 edge state. The curvature, on the one hand, enhances the coupling of the potential wells in Lattice-2. On the other hand, it induces a ring-shaped structure of the limit potential wells, that is, the larger the curvature, the closer the curved chain comes to forming a complete ring. It is worth figuring out which of the above two factors dominates for the enhancement of the L1 edge state. To this end, we choose a fixed curvature, 0.05 [see the example in Fig.~\ref{fig:3}(a)], to perform two additional calculations: 1) Increasing only the number of the potential wells to form a nearly closed ring shape but reserving the coupling of the lattices, and 2) slightly increasing only the size of the potential wells in Lattice-2 to increase its intra-lattice coupling but keeping the number of the potential wells fixed. The results are shown in Figs.~\ref{fig:4}(b,c), and the state in Fig.~\ref{fig:4}(a) is the original state, corresponding to the black point in Fig.~\ref{fig:3}(d), for comparison. It can be seen that more potential wells lead to the shift of the amplitude peaks to the left edge, but the whole structure can still be clearly seen [Fig.~\ref{fig:4}(b)]. Whereas, slightly increasing the diameter of the potential wells in Lattices-2 from 2.8~$\mu$m to 3~$\mu$m strongly pushes the strongest peak amplitude to the left edge, with the amplitude decreasing dramatically in the bulk wells [Fig.~\ref{fig:4}(c)]. This demonstrates that the increased intra-lattice coupling in Lattice-2 plays a crucial role in the enhancement of the L1 edge state, since a stronger intra-lattice coupling significantly deforms the higher-order modes in Lattice-2, as presented in Fig.~\ref{fig:1}(g). The intra-lattice coupling of Lattice-2 can be further strengthened by increasing the potential depth to obtain a more prominent edge state, as shown in Fig.~\ref{fig:4}(d), in which the potential wells have the same size as in Fig.~\ref{fig:4}(c) but a slightly larger depth. This is because the increased $V_2$ traps a stronger quadrupole amplitude in it; hence, it amplifies the difference of the inter- and intracell coupling. The potential depth of $V_2$ cannot be increased too much; otherwise, a much higher-order mode can be activated, whose deformation may be weakened and whose corresponding band gaps may become much narrower, disabling the topologically protected edge states in Lattice-1; see Figs.~\ref{fig:3}(d,e).

\section{Edge States in Nonlinear Regime}
As studied in \cite{schneider2024topological}, a repulsive nonlinearity that leads to an energy blueshift can lift specific bulk states into the band gap, forming edge states, i.e., nonlinearity-enhanced edge states. In Figs.~\ref{fig:3}(d,f), the curvature of a lattice pushes the corresponding bulk state to be localized at one of the edges. It is worth asking whether these states can be further pushed to form edge states. The bulk states in Fig.~\ref{fig:3}(b) at a curvature slightly larger than 0.055 show a similar profile to that in Fig.~\ref{fig:4}(c), raising the same question of whether such bulk states can be transitioned into edge states. To this end, we use a resonant,{spatially homogeneous} pump to excite the system into the nonlinear regime and study its impact on the existence of edge states. To obtain the stationary limit at a specific driving frequency $\omega$, in Eq.~\eqref{GP_psi}, the resonant pump is introduced as a continuous wave and has the form $E(\textbf{r},t)=E_0e^{-i\omega{t}}$ with a homogeneous amplitude $E_0$. Then, the corresponding time evolution can be efficiently solved by using a numerical solver~\cite{wingenbach2025phoenix}.

\begin{figure}[t]
\centering
{\includegraphics[width=1\linewidth]{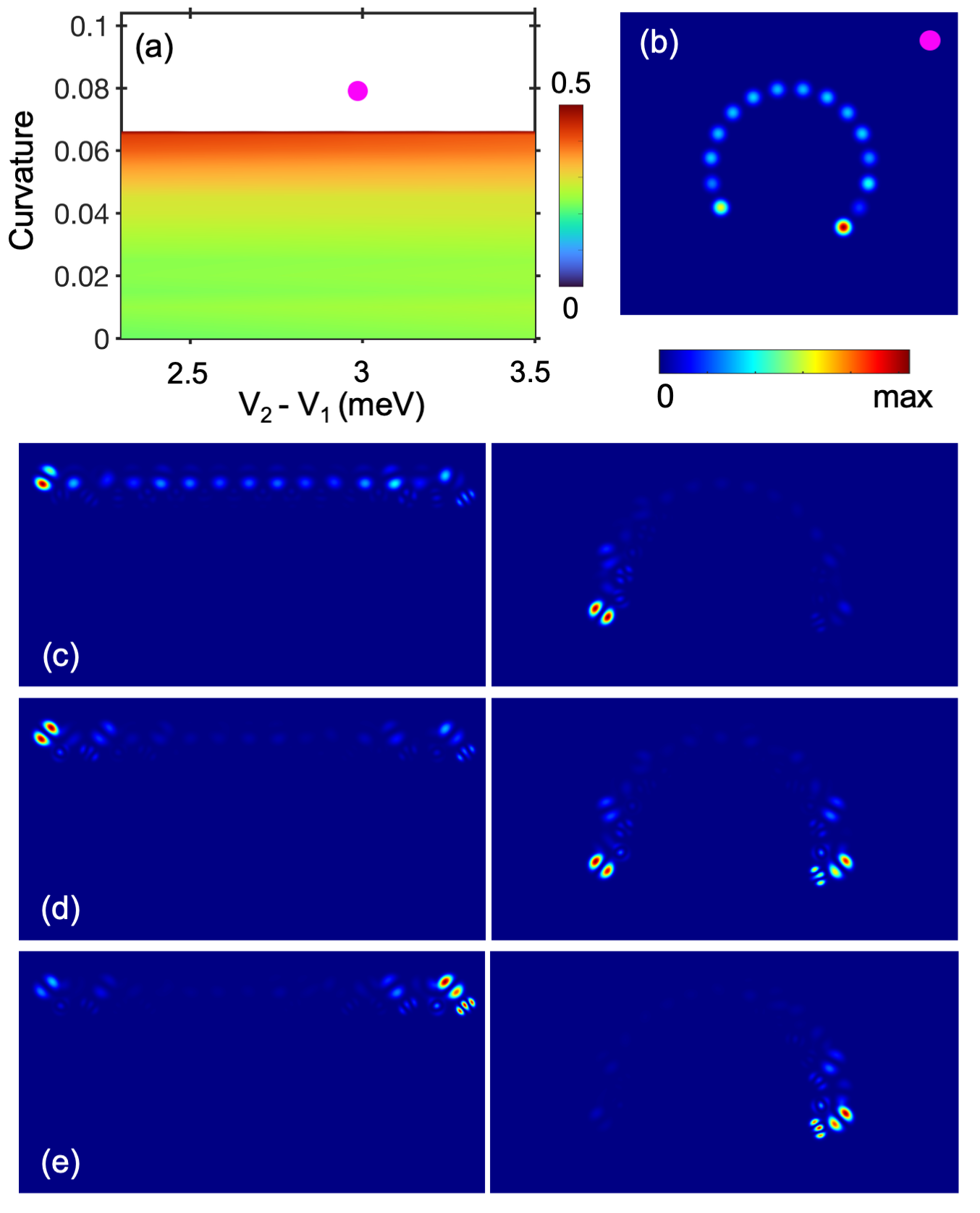}}
\caption{\textbf{Edge states in the nonlinear regime.} (a) Existence area of the R1 edge state in the nonlinear regime, i.e., under resonant excitation with $E_0=10.234~\mu$eV~$\mu$m$^{-1}$, based on the linear results in Fig.~\ref{fig:3}(b). (b) Density ($|\Psi|^2$, in $\mu$m$^{-2}$) distribution of an example bulk state, corresponding to the point in (a). (c-e) Density distributions of the coexisting L2 and R3 edge states at standard superlattices (left column), i.e., the curvature is 0 as shown in Fig.~\ref{fig:1}(c), and at curved superlattices (right column) with the curvature of 0.06, referring to the lattice structures in Fig.~\ref{fig:3}(a). The potential depths are (c) $V_2=7.9$~meV, (d) $V_2=8$~meV, and (e) $V_2=8.1$~meV. Here, the excitation amplitude is $E_0=66.522~\mu$eV~$\mu$m$^{-1}$. In all cases, $V_1=5$~meV.}
\label{fig:5}
\end{figure}

Under resonant excitation, the existence area of the R1 edge state in the curvature and potential depth domains can be found in Fig.~\ref{fig:5}(a). Compared with the result in the linear case in Fig.~\ref{fig:3}(b), the ratios at smaller curvatures are slightly increased (not blue anymore) because the finite loss rate perturbs the surrounding bulk states, as can be seen in Fig.~\ref{fig:1}(f), where the edge states are not well separated from the lower bulk states. However, the existence area of the R1 edge state is slightly extended to larger curvatures since the energy blueshift raises the edge states that already seep into the bulk states caused by the curvature, giving nonlinearity-induced enhancement.

Figures~\ref{fig:1}(f) and \ref{fig:2}(a) show that the edge states L2 and R3 can appear in the same gap and become nearly resonant around $V_2-V_1=3$~meV, guaranteeing their coexistence in the nonlinear regime. The bulk state in Fig.~\ref{fig:5}(b) shows that in the nonlinear regime, its profile is slightly different from those in the linear regime, cf. Fig.~\ref{fig:4}(c). That is, under excitation, both edges become dominant, which may favor the coexistence of the L2 and R3 edge states in the nonlinear regime. The evidence is presented in Fig.~\ref{fig:5}(d), in which the state in a straight chain (the left one) has a stronger L2 edge than the R3 edge. When the chain is bent to a larger curvature, 0.06, for instance, the R3 edge is enhanced, revealing a prominent coexistence of both states. If one of the two edge states is tuned to be much stronger than the other by, for example, slightly modulating the potential depth $V_2$, only one of them becomes favorable at larger curvatures [see the examples in Fig.~\ref{fig:5}(c,e)]. The results in Fig.~\ref{fig:5}(c-e) also guarantee edge state switching. The coexistence of other edge states may also be achieved under different system parameters.

\section{Conclusion}
To conclude, we have demonstrated topological edge states in zigzag polariton superlattices with imbalanced intra- and intercell couplings, which can be achieved by two mechanisms found in our work, i.e., the tunneling effect of the wavefunctions and the higher-order mode deformation. These mechanisms can be influenced by an additional degree of freedom, the curvature of the superlattices, which is introduced by arranging the lattices into curved shapes. In this case, the larger the curvature of the chain, the less prominent the edge states that originate from the tunneling effect. On the other hand, a curved chain with a larger intra-lattice coupling of the inner sub-lattice can strongly deform the higher-order modes, which increases the imbalance of the intra- and intercell coupling and consequently benefits the induced edge states. To enable the coexistence of the edge states in the same bandgap, coherent pumps can be applied to drive the density of the condensates into the nonlinear regime. Our design favors configurable topological edge states in more complex structures with respect to the introduced curvature.

Note that although we focus on microcavity polariton systems here, our model can be effortlessly extended to other physical systems, such as atomic condensates, nonlinear optics, and photonic crystals and waveguides. The curvature of the chains cannot be easily tuned in microcavity polariton systems as long as they are manufactured on top of the samples, but an optically imprinted potential provides an opportunity~\cite{pieczarka2021topological}. In addition, flexible materials and their optical properties have recently attracted significant attention~\cite{righini2021flexible}, which may enable tunable topological edge states and the state transition by, for example, bending or stretching the materials.

\section*{Acknoledgements}
This work was supported by the Deutsche Forschungsgemeinschaft (DFG) (No. 519608013 and No. 467358803), by the Paderborn Center for Parallel Computing, PC$^2$, and by the Semiconductor and Nano-Technology Center at Guangdong University of Technology.


%

\end{document}